\documentclass{aastex}
\input epsf

\newcommand{\msun}{\mbox{M$_{\odot}$}}

\newcommand{\vi}{\mbox{$V\!-\!I$}}
\newcommand{\bvz}{\mbox{$(B\!-\!V)_0$}}
\newcommand{\viz}{\mbox{$(V\!-\!I)_0$}}
\begin{document}
\title{Keck spectroscopy of young and old star clusters in NGC~1023
 \footnote{Based on data obtained at the W.M. Keck Observatory, which is 
 operated as a scientific partnership among the California Institute of 
 Technology, the University of California and the National Aeronautics and 
 Space Administration.}
}

\author{S{\o}ren S. Larsen and Jean P. Brodie
  \affil{UC Observatories / Lick Observatory, University of California,
         Santa Cruz, CA 95064, USA}
  \email{soeren@ucolick.org and brodie@ucolick.org}
}

\begin{abstract}
  We present Keck/LRIS spectra for 11 old globular clusters in the
S0-type galaxy NGC~1023 and 2 young blue clusters in the nearby companion
NGC~1023A.  Analysis of the spectra of 7 old clusters with
sufficient S/N shows generally good agreement between spectroscopic 
and previous photometric metallicity estimates, but the integrated
colors of two clusters are too blue for their spectroscopic metallicities.
Although the cluster ages are not well constrained, they are most likely
similar to those of Milky Way globular clusters and certainly older than
$\sim5$ Gyr.  The brightest GC in the sample shows enhanced cyanogen, 
possibly indicating an abundance anomaly similar to that observed in some 
M31 globular clusters.  

For the two blue clusters in NGC~1023A we estimate ages between 125 and 
500 Myr, based on their strong Balmer lines.  The total masses are about 
$8\times10^4$ M$_{\odot}$ and $6\times10^4$ M$_{\odot}$ for a Miller-Scalo 
IMF and $Z=0.004$, making these objects similar to the young populous 
clusters in the Large Magellanic Cloud. It is suggested that the two 
young clusters might have formed during a period of enhanced star formation 
activity in NGC~1023A, stimulated by a close encounter with NGC~1023.
\end{abstract}

\keywords{galaxies: star clusters ---
          galaxies: individual (NGC~1023)}

\section{Introduction}

  At a distance of around 10 Mpc \citep{cia91}, NGC~1023 is one of the 
nearest S0--type galaxies. It is included in the \emph{Atlas of peculiar 
galaxies} \citep{arp66} because of a small companion galaxy, NGC~1023A, 
which may have suffered a recent close encounter with NGC~1023 itself 
\citep{san84}. 
  The companion galaxy is quite noticably bluer than NGC~1023.
Its integrated broad-band colors, $U-B = 0.0$ and $B-V=0.7$ 
\citep{cap86} would imply a metallicity of [Fe/H] = $-1.7$ or less for 
an entirely old stellar population \citep{bc2001}, or the galaxy might have 
experienced recent star formation activity.  There are no traces of line 
emission in its spectrum \citep{cap86}, indicating no current star formation, 
but stars might have formed until a few 100 Myr ago when any residual gas was 
stripped during a close encounter with NGC~1023.  With an absolute $B$ 
magnitude of $M_B=-15.7$ \citep{cap86}, NGC~1023A is about one magnitude 
fainter than the Small Magellanic Cloud \citep{devau91}, but the ongoing 
star formation in the SMC presumably gives it a somewhat lower mass-to-light 
ratio than NGC~1023A so the two galaxies might have quite comparable masses.  

  Although globular clusters (GCs) are easily detectable at the distance of 
NGC~1023, its GC system has received surprisingly little attention. This might 
be related 
to the fact that NGC~1023 is located at a Galactic latitude of only $-19\deg$, 
causing serious contamination problems from foreground stars in ground-based 
studies. A recent study of GCs in NGC~1023 was carried out by \citet{lb00} 
(LB2000), who detected 221 cluster candidates with a clearly bimodal color 
distribution on HST / WFPC2 images. Among the clusters identified by LB2000 
were a number of red objects, which were generally fainter and had larger 
effective radii than normal GCs.  In this paper we discuss new spectroscopic 
data for normal compact globular clusters in NGC~1023 and for two young 
clusters in the companion galaxy, NGC~1023A.  

\section{Observations}

  Objects to be observed with the LRIS spectrograph \citep{oke95} on the 
Keck I telescope were picked from the list of GC candidates identified
by LB2000. Additional objects to fill up the slitmask, including two
blue objects in NGC~1023A, were selected 
from ground-based CCD images from the 1.2 m telescope at F.\ L.\ Whipple 
observatory on Mount Hopkins, Arizona, kindly provided by P.\ Barmby. 
The Keck spectra were obtained in February 2000, using 1\arcsec\ slitlets
and a 600 lines/mm grating blazed at 5000 \AA .  This provides a
spectral resolution of about 6 \AA\ and a scale of 1.28\AA\ per pixel.
The spectra typically covered the region 3900 \AA\ -- 6300 \AA,
but the limits varied by a few 100 \AA\ from one spectrum to another.
Altogether the slitmask included 20 science objects, of which 10 were 
selected from the LB2000 WFPC2 images.  The total integration time was 9300 
sec, split into 6 individual exposures. Flux standards from \citet{mas88} 
(PG0838$+$546 and PG0939$+$262) and 4 radial velocity standard stars (HR1805, 
HR3905, HD37251 and HD52071) were observed through a 1\arcsec\ longslit. 
HR1805 and HR3905 were also observed by \citet{w94} in the Lick/IDS system,
making them suitable for a consistency check of our own spectroscopy.

  The multislit spectra were extracted from the individual images using
the custom-designed REDUX package, written by A.\ Phillips. 
REDUX automatically detects each object spectrum and subtracts the sky
background, based on information in the files that were originally used 
for designing the slitmasks.  The wavelength solution was obtained by REDUX 
from exposures of arc spectra through a special calibration slitmask, with 
corrections to the wavelength zero-point from night sky lines. Co-addition
of the individual integrations was also performed by REDUX. The flux-- and 
radial velocity standards were reduced using standard tools in the IRAF 
SPECRED package\footnote{IRAF is distributed by the National Optical 
Astronomical Observatories, which are operated by the Association of 
Universities for Research in Astronomy, Inc.~under contract with the National 
Science Foundation}, and the flux calibrations were then applied to the 
multislit spectra.  
Radial velocities were obtained by cross-correlation of 
the object spectra with the spectra of the four radial velocity standards, 
using the FXCOR task in the RV package in IRAF.  

  Basic data for the science objects are listed in Table~\ref{tab:spec}.
The first column in the table is an object identifier and the second column
indicates whether the object was selected from the ground-based or
WFPC2 data. The following columns list the 2000.0 coordinates and
$V,I$ photometry (uncorrected for reddening). 
Column 7 gives the mean heliocentric radial velocity from the four individual
measurements and the standard error on the mean value.
Finally, column 8 lists the 
signal-to-noise per pixel in the dispersion direction of each spectrum in 
the region 4900\AA -- 5100\AA .

  The radial velocity of NGC~1023 is listed as $601\pm11$ km/sec in the 
RC3 catalogue, while \citet{cap86} find a somewhat higher radial velocity of
$742\pm30$ km s$^{-1}$ for NGC~1023A.
Many of the objects selected on the ground-based images were near the 
detection limit (as is evident from the large photometric 
errors) and some are probably spurious detections (n1023-7, n1023-12, 
n1023-23). Others did not reveal a clear cross-correlation signal anywhere 
near the NGC~1023 radial velocity (n1023-18, n1023-20) and are presumably 
background galaxies. n1023-18 shows broad emission lines characteristic
of quasars, corresponding to the CIV and CIII$]$ lines at rest wavelengths
1549 \AA\ and 1909 \AA, redshifted to $z=2.15$ (Fig.~\ref{fig:spectra}). 
n1023-24 is probably a foreground star.  Two objects are good GC candidates 
(n1023-13 and n1023-25), and we call particular attention to object 
n1023-13 (Fig.~\ref{fig:spectra}) whose apparent $V$ band magnitude 
corresponds to an absolute magnitude of $M_V = -10.8$ at the distance of 
NGC~1023.  This cluster, discussed in detail in \citet{lar01}, is about half 
a magnitude brighter than the brightest globular cluster in the Milky Way, 
$\omega$ Centauri \citep{har96}.  The two blue objects in NGC~1023A, n1023-3 
and n1023-4, have strong Balmer lines, indicating the presence of A-type 
stars, and radial velocities consistent with the NGC~1023 system and with 
NGC~1023A in particular.  Of the HST-selected objects, n1023-15 is presumably 
a background galaxy, but the remaining 9 objects are good GC candidates.  
Thus, our data include 11 old globular clusters in NGC~1023 and 2 young 
objects in NGC~1023A.  The mean radial velocity of the whole sample is 597 
km/s with a dispersion of $\pm154$ km/s, or $573\pm155$ km/s if the two young 
clusters are excluded, in nice agreement with the radial velocity of NGC~1023 
quoted above.

\section{Spectrophotometric analysis of abundances and ages}

  Although the metallicities of globular clusters are generally believed
to correlate with their broad-band colors, a spectroscopic analysis provides 
more detailed information on possible anomalies and/or age differences. An 
empirical calibration of various spectrophotometric indices as a function of
metallicity was published by \citet[hereafter BH90]{bh90}, and more 
recently Lick/IDS indices for stellar populations of various ages and 
metallicities have been modeled by e.g.\ \citet{wor94}, \citet{bc2001}, C
.\ Maraston and others.  

  The BH90 metallicity calibration is based on a number of 
metallicity-sensitive spectrophotometric indices, each of which provides an 
estimate of the metallicity. A weighted mean of these individual measurements 
then provides the final spectroscopic metallicity estimate of the cluster,
where the appropriate weights are determined from the random errors on the
index measurements as well as from the inherent uncertainty in the metallicity
derived from a given index. The latter contribution is given by the
``figure of merit'' for each index, as defined by BH90.  Even though we will 
often label these estimates by [Fe/H], it should be emphasized that the 
metallicity calibrators really measure a variety of different elements, 
including C, N, Mg and Fe. In our treatment of the errors we adopt one
modification from the BH90 prescription, whose equation (22) gives the
weighted dispersion of the individual [Fe/H] measuremens rather than
the uncertainty on the weighted mean. Thus, we multiply by a factor
$(\sum W_j -1)^{-1/2}$ so that our error on the combined [Fe/H] 
estimate, $\sigma_W$, becomes
\begin{equation}
  \sigma_W = \left[
             \frac{\sum W_j(\left[{\rm Fe/H}\right]_I -
	                    \left[{\rm Fe/H}\right]_W)^2}
                  {\left(\sum W_j\right) \left(\sum W_j -1\right)}
             \right]^{1/2}
\end{equation}
where $W_j$ are the weights (normalized to the range $0\leq W_j \leq 1$) and 
[Fe/H]$_I$ and [Fe/H]$_W$ are the individual and weighted mean metallicity 
estimates.

  Spectrophotometric indices are generally based on a central bandpass 
covering the feature of interest and two bandpasses on either side of the 
feature bandpass which define a ``pseudo-continuum''. Several variants
of the exact prescription exist in the literature, mainly differing in the 
way the fluxes in the continuum bandpasses are interpolated to the
feature bandpass \citep[BH90]{bur84,w94}. In practice, the definitions
by \citet{bur84} and \citet{w94} usually give virtually identical results, 
while indices computed according to BH90 can be slightly different, especially 
if the continuum passbands are located asymmetrically with respect to the 
feature bandpass and the spectrum has a significant slope. 

\subsection{Metallicities}

  We derived spectroscopic metallicities for old globular clusters with 
S/N$>$5 using the calibration in BH90.  Indices were measured
according to their prescription after correction for radial velocities as
listed in Table~\ref{tab:spec}. If instead the \citet{bur84} or 
\citet{w94} definitions were used, we found a slight decrease in the [Fe/H] 
values by about 0.02 dex, with most of the difference coming from the Fe5270 
index.  Thus, it is clear that the exact prescription used for measuring the
indices has only a very weak effect on the resulting metallicities.
As a check, we also compared our index measuring software with the SBANDS 
task in IRAF (which uses the Burstein et al.\ definition) and got identical 
values when using that same definition.

  Each of the individual metallicity estimates, as well as the mean values, 
are given in Table~\ref{tab:metal} for the 7 old clusters with $S/N>5$.
Of these clusters, 6 belong to the ``blue'' peak in the color
distribution as defined by LB2000 [$\viz < 1.05$] and one (n1023-25) 
belongs to the ``red'' peak.  Because none of the spectra reached sufficiently 
far into the blue, the BH90
$\Delta$ index could not be measured for any of the clusters and has
therefore not been included in the Table. Furthermore, we have excluded
the NaD index because of the strong NaD skylines, which at best add large
amounts of noise to measurements of this index and at worst may cause 
systematic errors.  For comparison, we also list metallicities derived from 
the reddening-corrected \vi\ colors, using the calibration in \citet{kis98}
and adopting a foreground extinction of $A_B=0.26$ \citep{sch98}.
The uncertainties on these photometric metallicities only include the 
random errors due to the photometry, while no systematic errors in the
calibrations etc.\ have been taken into account.

  Although most of the individual metallicity estimators are compatible with 
the average values within the errors for any given cluster, it may be worth 
noting that the measurements based on the Fe5270 index tend to give higher 
metallicities than the other indices. Typically, the [Fe/H] values based on 
the Fe5270 index fall about 0.5 dex above the mean values.
In Figure~\ref{fig:fehcmp2} we compare the spectroscopic and photometric 
metallicities.  There is generally fair agreement between the two metallicity 
estimates, although we note the presence of two outliers (n1023-14 and 
n1023-16) for which the spectroscopic measurements give significantly
higher metallicities than the photometric ones.  The reason for this is not 
clear, but from Table~\ref{tab:metal} the discrepancy does not appear to
be driven by any individual index, except for the high metallicity from 
the Fe5270 index which is common to all clusters. In fact, \emph{all} of the
individual metallicity estimators give higher values than the photometric 
estimate for these two objects.  We also measured ``pseudo-colors'' for the 
clusters on the spectra at 4700 \AA\ and 5700 \AA\ in 500 \AA\ windows 
(Table~\ref{tab:ctest}), confirming that n1023-14 and n1023-16 have rather 
blue colors for their metallicities. 

  The data used by BH90 came from a variety of instruments with
different spectral resolutions, ranging from about 5\AA\ (AAT and MMT) to
12\AA\ (Lick).  To test how the results depend on the spectral resolution, 
we smoothed our data with an 8 pixels (10\AA ) wide Gaussian kernel to 
simulate the 12\AA\ resolution of the lowest resolution spectra used by BH90 
and redid the metallicity measurements.  Such smoothing typically decreased 
the metallicities by $\sim0.1$ dex. Since only a fraction of the BH90 data 
were observed at 12 \AA\ resolution and the rest were observed at a 
resolution quite similar to ours, systematic errors on our mean metallicity 
estimates due to resolution effects should be smaller than 0.1 dex. 

  \citet{bur84} and \citet{bh91} noted that many GCs in M31 have enhanced 
CNR indices compared to Galactic globulars of the same metallicity.  
Table~\ref{tab:indices_bh} lists a number of indices for the 7 brightest
old GCs in NGC~1023, and in Fig.~\ref{fig:fehcmp3} we show the CNR indices 
versus [Fe/H]. The polygon represents the approximate region of 
the diagram populated by Galactic globulars, according to \citet{bh91}.  
We note that two clusters (n1023-13 and n1023-14) do seem to have enhanced 
CNR indices, although this may only be significant for n1023-13 due to the 
larger errors on the data for n1023-14.  For the remaining spectra the errors 
are too large to put strong constraints on any anomalies although most of them 
seem to be compatible with the Milky Way data.

\subsection{Comparison with population synthesis models}


  Age differences between clusters 
would manifest themselves most strongly in the Balmer lines. Beyond
a few hundred Myrs the strength of the Balmer lines decreases as a
function of age, as the main sequence turn-off shifts towards cooler 
temperatures.  However, it is important to note that the appearance of 
blue horizontal-branch stars in metal-poor populations around 10 Gyr 
can reverse this trend, potentially making the Balmer lines less useful as 
age indicators in very old, metal-poor populations \citep{mt00}. Also,
many population synthesis models suffer from uncertainties in the
modeling of AGB stars which can have a significant effect on the Balmer
indices \citep{schia02}. With those precautions in mind, we 
compare compare our data with model predictions from \citet{bc2001} (BC2001) 
in the following.

  There are a few instrumental effects that need to be considered when
comparing our measurements with the BC2001 models, which use
line indices based on the \citet{w94} fitting functions. First, the Lick/IDS
spectra were not flux calibrated and second, our $\sim6$ \AA\ resolution
is higher than that of the Lick/IDS spectra.  In order to test how 
sensitive our measured indices are to these effects, Table~\ref{tab:stdcmp}
lists measurements of a selection of Lick/IDS indices on  flux-calibrated
spectra of the two Lick standards HR1805 and HR3905, as well as on unfluxed 
spectra and spectra smoothed with a 5 pixels wide Gaussian kernel to simulate 
the lower Lick/IDS resolution.  Generally, flux calibration only affects the 
equivalent widths by a few times 0.01\AA , in most cases much smaller than the 
random measurement errors.  This has also been documented by 
other authors \citep[e.g.][]{fab85,kis98}, but see \citet{bea00} for some
cautionary notes.  However, the smoothing does have a significant effect on 
the results in some cases.  The EWs measured on the unsmoothed data are
generally too high by several times 0.1 \AA\ compared to the Worthey et al.\
standard values, while smoothing of the data clearly results in better 
agreement.  We therefore decided to smooth all the science spectra with a 5 
pixels Gaussian kernel before measuring Lick / IDS indices.

  Selected Lick/IDS indices for the NGC~1023 clusters are listed in 
Table~\ref{tab:indices_gw}.  The last row gives the mean indices for all of 
the 6 metal-poor old clusters, i.e.\ excluding n1023-3, n1023-4 and n1023-25.  
In Figure~\ref{fig:hb_mg2} and \ref{fig:hb_fe5270} we compare the Fe5270, Mg2 
and H$\beta$ measurements in Table~\ref{tab:indices_gw} with the BC2001 models.
The mean values from Table~\ref{tab:indices_gw} are indicated with $*$ symbols. 
  From Figure~\ref{fig:hb_mg2} and \ref{fig:hb_fe5270}, most clusters are
at least several Gyrs old, except for n1023-3 and n1023-4 whose very
strong Balmer lines imply ages less than 500 Myr (see Section~\ref{sec:cc}).
The combined datapoint falls near the 10/15 Gyr isochrones in both 
plots, and between the [Fe/H] = $-0.7$ and $-1.7$ lines, in agreement with 
the metallicities derived from the BH90 calibration.  However, the separation 
between the 5, 10 and 15 Gyr isochrones in these diagrams is comparable to 
the errors even for the combined data and the theoretical 10 Gyr and 15 Gyr 
lines actually \emph{cross each other} at [Fe/H] $\sim -1$.  
Both Fig.~\ref{fig:hb_mg2} and Fig.~\ref{fig:hb_fe5270} formally indicate
an age of $\sim 1$ Gyr and above solar metallicity for the faint cluster 
n1023-25, but the error bars are large and it seems premature to conclude
that this cluster is truly young.  It would be interesting to 
obtain high S/N spectra of additional metal-rich GCs in NGC~1023 and look
for age differences between metal-poor and metal-rich clusters.

  To summarize, the spectroscopy generally confirms the photometric
metallicity estimates, but we find that two clusters have significantly
higher spectroscopic metallicities than  would be expected from their
integrated colors.  We note that the bright cluster n1023-13 shows
a possible enhancement in the CNR index, similar to that observed in some
M31 clusters.  Most of the GCs appear to be old ($\ga$ 5 Gyr) and have 
ages consistent with those of Milky Way globulars, but the ages are not 
well constrained.

\section{The two blue objects in NGC~1023A}
\label{sec:cc}

  \citet{dav84} noted a number of faint point sources in NGC~1023A, near 
the detection limit on their photographic plates, and suggested that these 
might be supergiant stars. The ground-based CCD images from which we selected 
our GC candidates for Keck spectroscopy clearly showed two blue sources
superimposed on NGC~1023A, which we decided to include in our sample.

  The spectra of these two objects, n1023-3 and n1023-4 turned out to
be nearly identical and only one of them is shown in Figure~\ref{fig:spectra}.
The most striking features of the spectra are the very strong Balmer lines,
an obvious indication that these are young objects dominated by
A-type stars, and thus with ages of no more than a few times $10^8$
years. The radial velocities are compatible with NGC~1023 membership,
and actually favor association with NGC~1023A. After correction for reddening, 
the ground-based photometry indicates
$\bvz = 0.25\pm0.06$ and $0.36\pm0.04$, 
$\viz = 0.62\pm0.08$ and $0.60\pm0.04$, and 
$M_V = -9.25\pm0.05$ and $-9.54\pm0.03$ 
for the two objects.
Only stars with masses above 35 \msun\ reach such high luminosities,
and have lifetimes less than 5 Myr \citep{ber94,gir00}. Star formation activity
that recent would still be easily detectable as H$\alpha$ emission,
which is clearly incompatible with the observations -- notwithstanding
the fact that it would be highly unlikely to find two isolated stars 
that massive. Thus, the two blue ``central condensations'' in NGC~1023 are 
most likely young star clusters. Unfortunately, they are not included in 
any HST images and the available ground-based images have insufficient 
resolution and S/N to tell whether or not they are spatially extended.

\subsection{Spectroscopic ages}

  Fig.~\ref{fig:hb_cc_gw} shows BC2001 model predictions for the 
evolution of the Lick/IDS H$\beta$ index as a function of age for three
different metallicities: $Z=0.02$ (solar), $Z=0.008$ and $Z=0.004$. The 
plot also includes model predictions by C.\ Maraston (private comm.) for 
solar metallicity.  The Maraston models are also available for sub-solar 
metallicities, but these have been omitted here for clarity.  Equivalent 
widths of the H$\beta$ line measured for the two blue objects in NGC~1023A are 
indicated by the hatched areas.  From Figure~\ref{fig:hb_mg2} and 
Figure~\ref{fig:hb_fe5270} the two objects appear to have subsolar 
metallicities, as expected if they formed in a stellar environment similar to 
that of the SMC, but the measurement and model uncertainties are too large to 
provide strong constraints on the exact metallicities.

  As can be seen from Fig.~\ref{fig:hb_cc_gw}, age determinations based on 
Balmer lines alone are ambiguous, because the predicted equivalent
widths reach a maximum at a few times $10^8$ years.
For solar metallicity ($Z=0.02$), comparison with the BC2001 models suggests 
cluster ages around 125 Myr or around 300 Myr, but for $Z=0.004$ (similar to 
that of the young stellar population in the SMC) the ages 
could be as high as $\sim500$ Myr.  The best match to the Maraston models 
is obtained at an age around 200--300 Myr.

The \citet{w94} fitting functions, used by both the BC2001 and the Maraston
models, were tailored primarily for studies of old stellar populations in 
early-type galaxies, and may produce less reliable results for younger stellar 
populations.  
  An alternative set of broader indices, tailored for young stellar 
populations, was introduced by \citet{bro98}. We hereafter refer to these
indices as B98 indices.  In Fig.~\ref{fig:ccspec} we compare B98 
(H$\beta$, H$\gamma$, H$\delta$) equivalent widths with model predictions
based on the spectral energy distributions (SEDs) supplied 
with the \citet{bc2001} models. B98 indices computed by C.\ Maraston 
(private comm.) produce essentially the same resuls.  Figure~\ref{fig:ccspec} 
also shows the model predictions for $B-V$ colors and the ground-based 
measurements.  

  Again, Figure~\ref{fig:ccspec} shows that both clusters are young 
objects with ages on the order of 300 Myr, but note that nearly all the
Balmer line indices are \emph{stronger} than the maximum values predicted by 
the models. A similar effect was observed by \citet{bro98} for young clusters 
in NGC~1275. They suggested that the unusually strong Balmer lines might be 
due to a skewed or truncated stellar IMF, and found that the observed Balmer 
line EWs were better reproduced by models with a lack of stars below 
$\sim 2 \msun$.  Although \citet{bro98} only adopted the truncated IMF models 
as the best available approximation to a flatter IMF, not necessarily implying 
a complete lack of low-mass stars in the real clusters, it is clear that
the IMF would have to be dramatically different from the ``standard''
Salpeter slope in order to produce significant changes in the observed
EWs of spectral lines. Using BC2001 models computed for a Miller-Scalo IMF
instead of Salpeter, for example, has virtually no impact on the predicted EWs.

  In conclusion, the integrated spectra of young stellar populations clearly
need to be better understood.  With current models we cannot constrain the 
age of the two blue objects in NGC~1023A to better than somewhere between 125 
and 500 Myr. Part of the uncertainty is due to the behavior of Balmer line
strengths as a function of age, but it is also worth noting that different
models and index definitions can provide significantly different results.
However, we note that the estimated cluster ages are compatible with 
the clusters having formed during the last close encounter between NGC~1023 
and NGC~1023A.

\subsection{Masses}

  Because of the uncertain ages and other factors, estimates of the total 
cluster masses are also inherently uncertain. However, we can make some
crude estimates: For a Miller-Scalo IMF and an age of 300 Myr, the BC2001
models predict a luminosity per solar mass of $M_V(1 \msun) = 2.74$ 
for $Z=0.004$. For $Z=0.02$ the models predict $M_V(1 \msun) = 2.95$. If 
we adopt an uncertainty 
of $\pm100$ Myr for the ages (which may be somewhat optimistic, cf.\ the 
discussion in the previous section) then the corresponding uncertainty on the 
$V$ band M/L ratio is about $\pm0.3$ mag. Considering the uncertainty on
metallicity, a more realistic error estimate on the M/L ratio
may be at least 0.5 mag.  With absolute $V$ band magnitudes 
of $-9.54$ and $-9.25$ the masses of the two clusters then become 
$8.2\times10^4\,\msun$ and $6.2\times10^4\,\msun$, with an uncertainty
of about $\pm50\%$.  For a Salpeter IMF extending 
down to 0.1\msun\ the masses are roughly a factor of two higher than those 
based on the Miller-Scalo IMF, i.e.\ $1.8\times10^5\,\msun$ and 
$1.3\times10^5\,\msun$ for the two clusters. For a truncated IMF 
the masses would be even smaller than for the Miller-Scalo IMF. However, it 
is clear that the two blue clusters in NGC~1023A are fairly massive, and 
they may be quite similar to young ``populous'' clusters such as
NGC~1866 in the LMC \citep{hod61}.

\subsection{Are the young blue clusters in NGC~1023A related to the
  ``faint fuzzies'' in NGC~1023?}

  LB2000 discovered a number of peculiar faint, extended objects in NGC~1023, 
thought to be stellar clusters associated with the disk of
the galaxy. These clusters have about the same \vi\ color as the metal-rich
``normal'' globular clusters in NGC~1023, but are generally fainter and
have larger sizes. Most of them have effective radii of
$\sim10$ pc and absolute visual magnitudes fainter than $M_V=-7$. While
their origin remains unknown, one possibility is that they might have
been accreted from dwarf galaxies similar to NGC~1023A. Some support for
this idea is provided by the presence of an HI-ring around NGC~1023, 
hinting at encounters in the system earlier in its history \citep{cap86}.
It is thus natural to ask how the two blue objects in NGC~1023A will 
evolve with time.

  Assuming that the two blue objects currently have ages of 300 Myr, they 
will fade by $\sim1$ magnitude by the time they are 1 Gyr old and by
about 3.5 mag when they reach an age of 10 Gyr. Thus, they will be in 
the magnitude range spanned by the ``faint fuzzies'' several Gyr from
now, provided they remain bound. This by itself, of course, does not
imply that the faint extended clusters in NGC~1023 originated in 
dwarf galaxies similar to NGC~1023A, but it would be highly desirable
to check the sizes of the two blue clusters with high-resolution imaging.


\section{Summary and conclusions}

  We have presented Keck spectroscopy for 11 old globular clusters in the 
nearby S0-type galaxy NGC~1023 and 2 young clusters in its companion galaxy, 
NGC~1023A.  Out of 10 objects from \citet{lb00}, 9 are confirmed to be 
globular clusters in NGC~1023 and in addition we have identified 2 new 
globular clusters, one of which is very luminous with $M_V\approx-10.8$.  Most 
of the spectra have rather poor S/N but 7 are good enough to allow analysis 
of metallicities. Two of these clusters have higher
metallicities than predicted by their broad-band colors.  We have looked for 
abundance anomalies and find that most clusters show no detectable anomalies, 
although the bright cluster n1023-13 may show
cyanogen enhancements similar to those observed in M31 globular clusters 
\citep{bur84,bh91}. Most of the clusters appear to have high ages consistent
with Milky Way globulars, but the errors are large and it would be highly 
desirable to check possible age differences between the metal-poor and 
metal-rich populations with more spectra of metal-rich GCs in NGC~1023.

  Two blue objects in NGC~1023A are young star clusters with ages 
around 300 Myr, but with a possible range between 125 and 500 Myr.  Their 
integrated $V$ band magnitudes are $M_V = -9.6$ and $M_V = -9.3$.  The total 
masses are about $8\times10^4\,\msun$ and 
$6\times10^4\,\msun$ for a Miller-Scalo IMF and SMC--like metallicity, 
making these clusters similar to young populous clusters in the LMC such 
as NGC~1866. Since NGC~1023 and NGC~1023A may have undergone a close encounter 
at about the same time the two blue clusters in NGC~1023A were formed, 
it is tempting to speculate that the clusters formed during a period of 
enhanced star formation activity, stimulated by the encounter.  

\acknowledgments

We are grateful to Pauline Barmby for providing us with her ground-based
CCD images of NGC~1023 and to Markus Kissler-Patig for helpful comments on 
an earlier version of this manuscript. We also thank John Huchra for
useful discussions and the anonymous referee for comments which helped
improve the manuscript.  This work was supported by National Science 
Foundation grant number AST9900732.

\newpage

\begin{minipage}{15cm}
\epsfxsize=7cm
\epsfbox{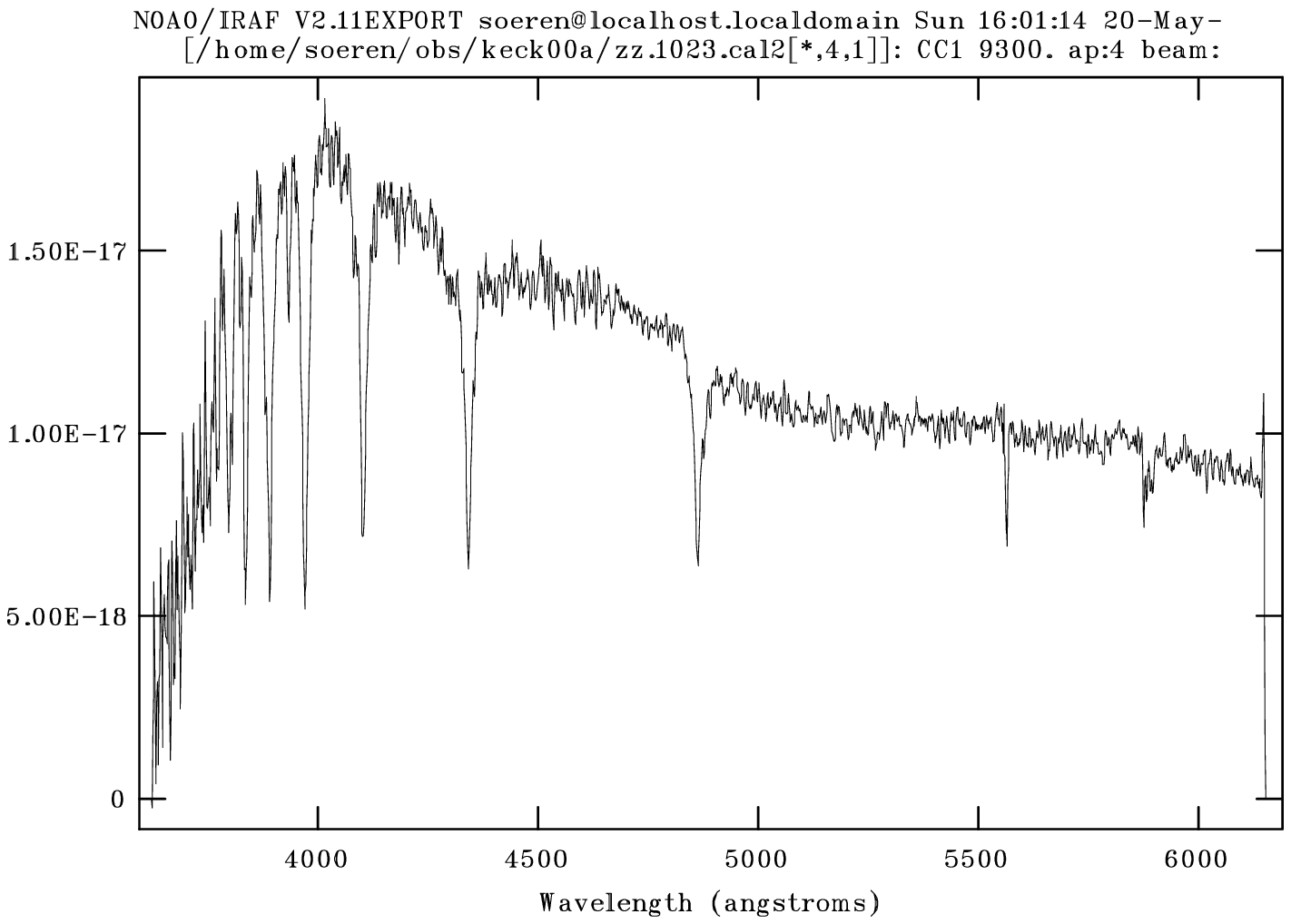}
\epsfxsize=7cm
\epsfbox{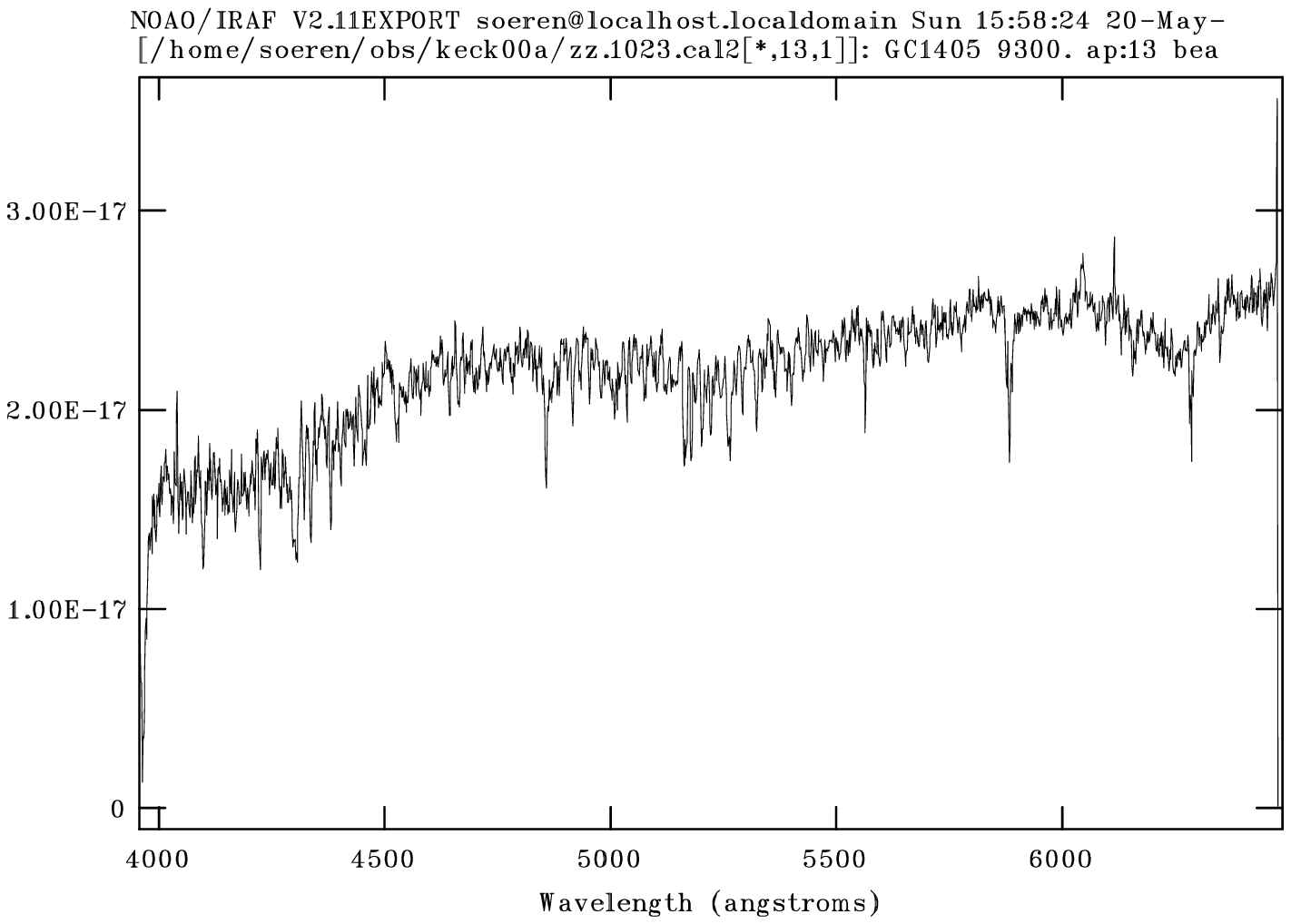}
\epsfxsize=7cm
\epsfbox{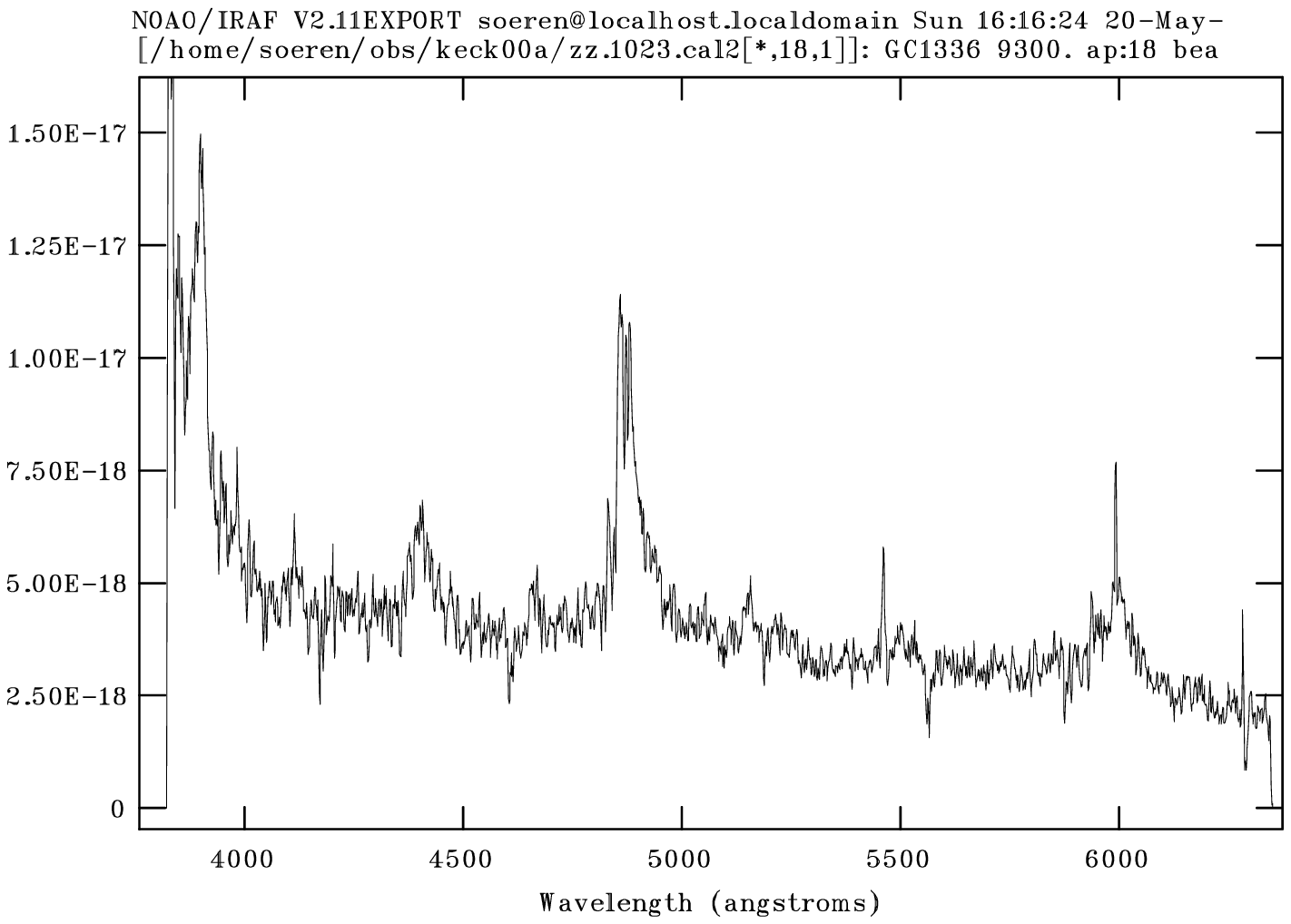}
\end{minipage}
\figcaption[Larsen.fig1a.ps,Larsen.fig1b.ps,Larsen.fig1c.ps]{\label{fig:spectra}
  (a) One of the blue objects in NGC~1023A (n1023-4). Note the very strong
Balmer lines.  (b) The bright globular cluster (n1023-13). (c) Object 
n1023-18, showing broad emission lines corresponding to CIV (1549 \AA ) 
and CIII$]$ (1909 \AA ) redshifted to $z=2.15$ and thus presumably a quasar.  
All spectra have been smoothed with a 3 pixels wide boxcar filter.
}

\newpage

\epsfxsize=14cm
\epsfbox{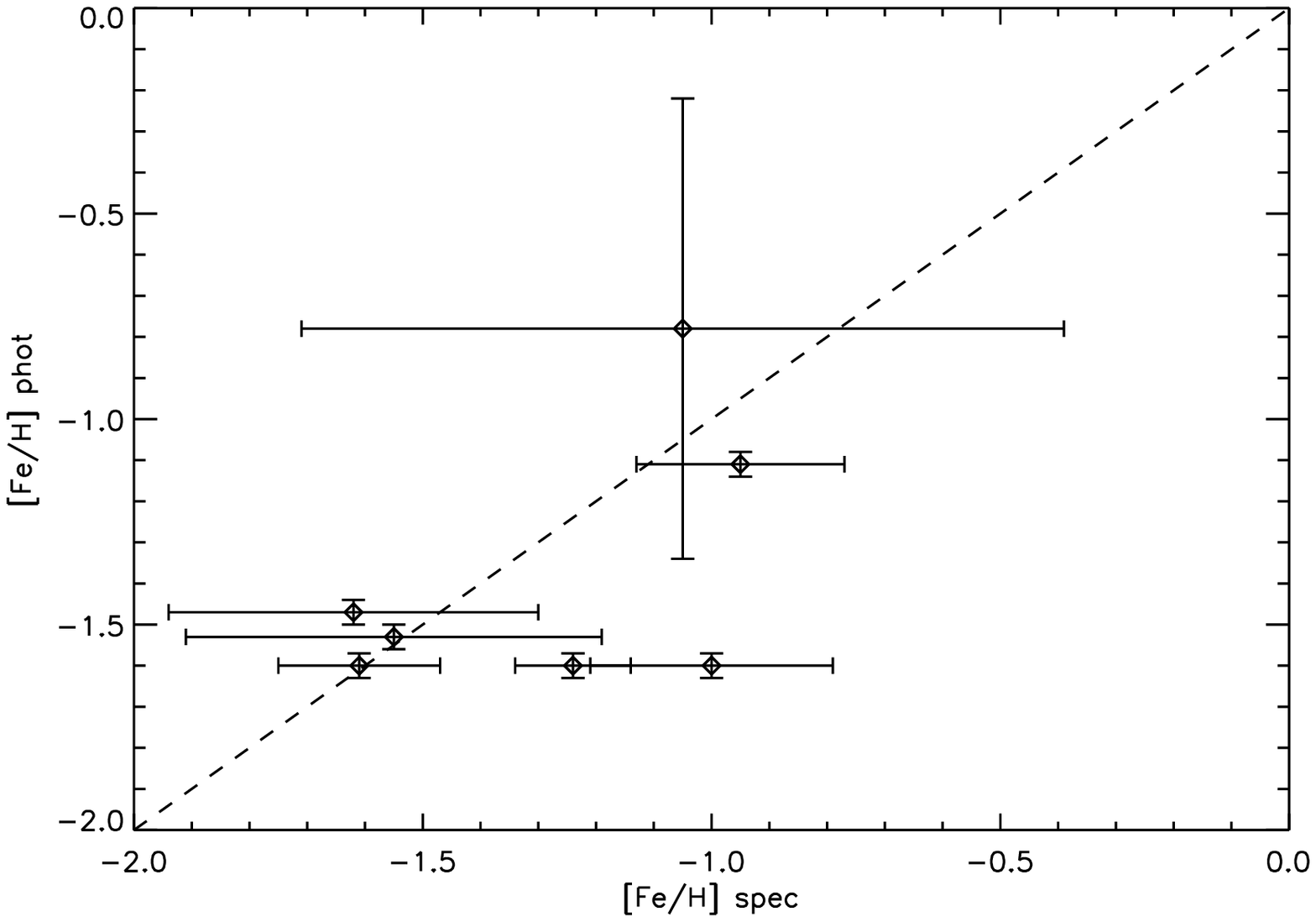}
\figcaption[Larsen.fig2.ps]{\label{fig:fehcmp2}Comparison of average 
spectroscopic versus photometric metallicities. The dashed line represents 
a 1:1 relation.}

\newpage

\epsfxsize=14cm
\epsfbox{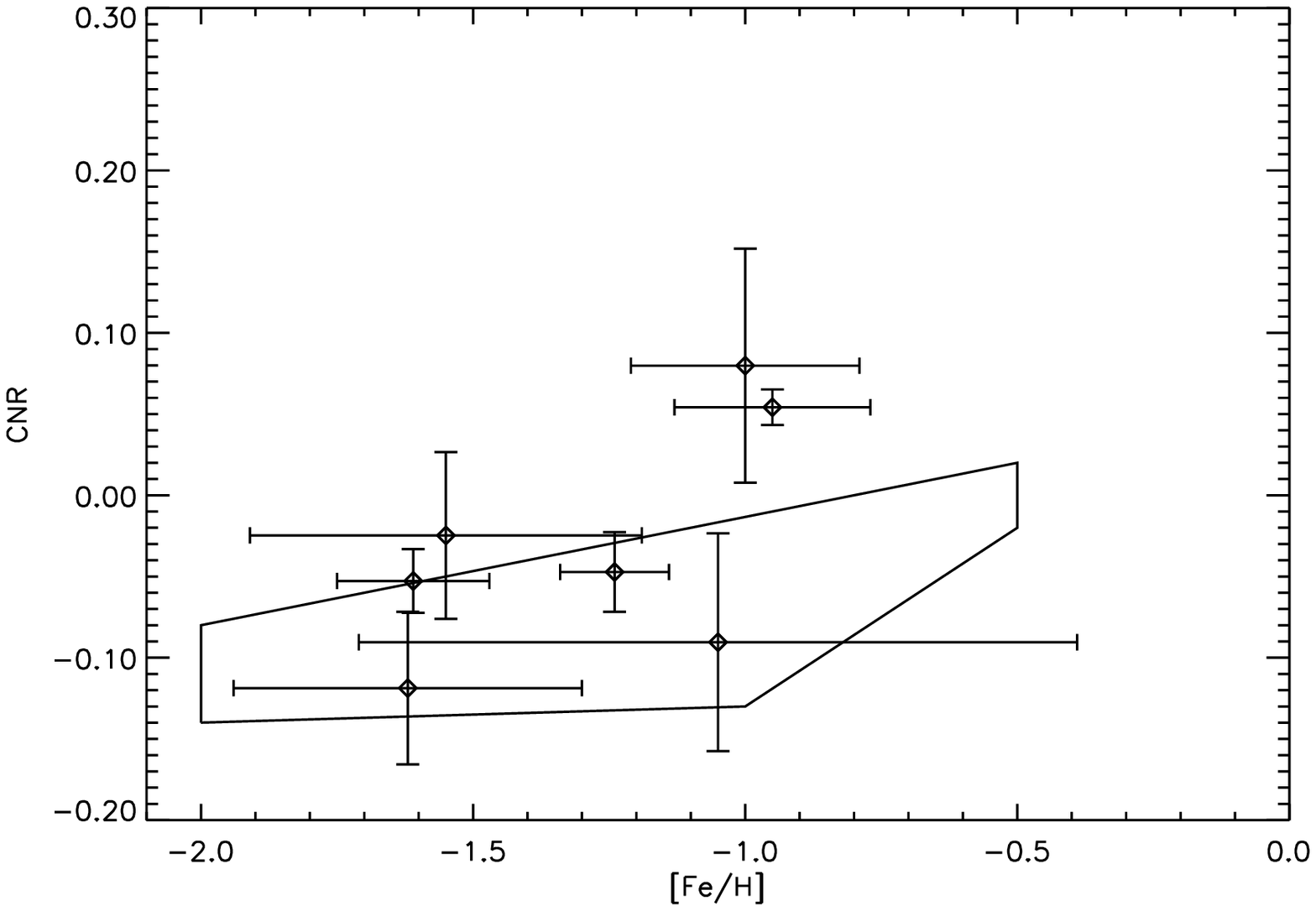}
\figcaption[Larsen.fig3.ps]{\label{fig:fehcmp3}
CNR indices as a function of metallicity.  The polygons outline the part of 
the diagram populated by Milky Way globulars (from Brodie \& Huchra 1991). 
}
\newpage

\epsfxsize=14cm
\epsfbox{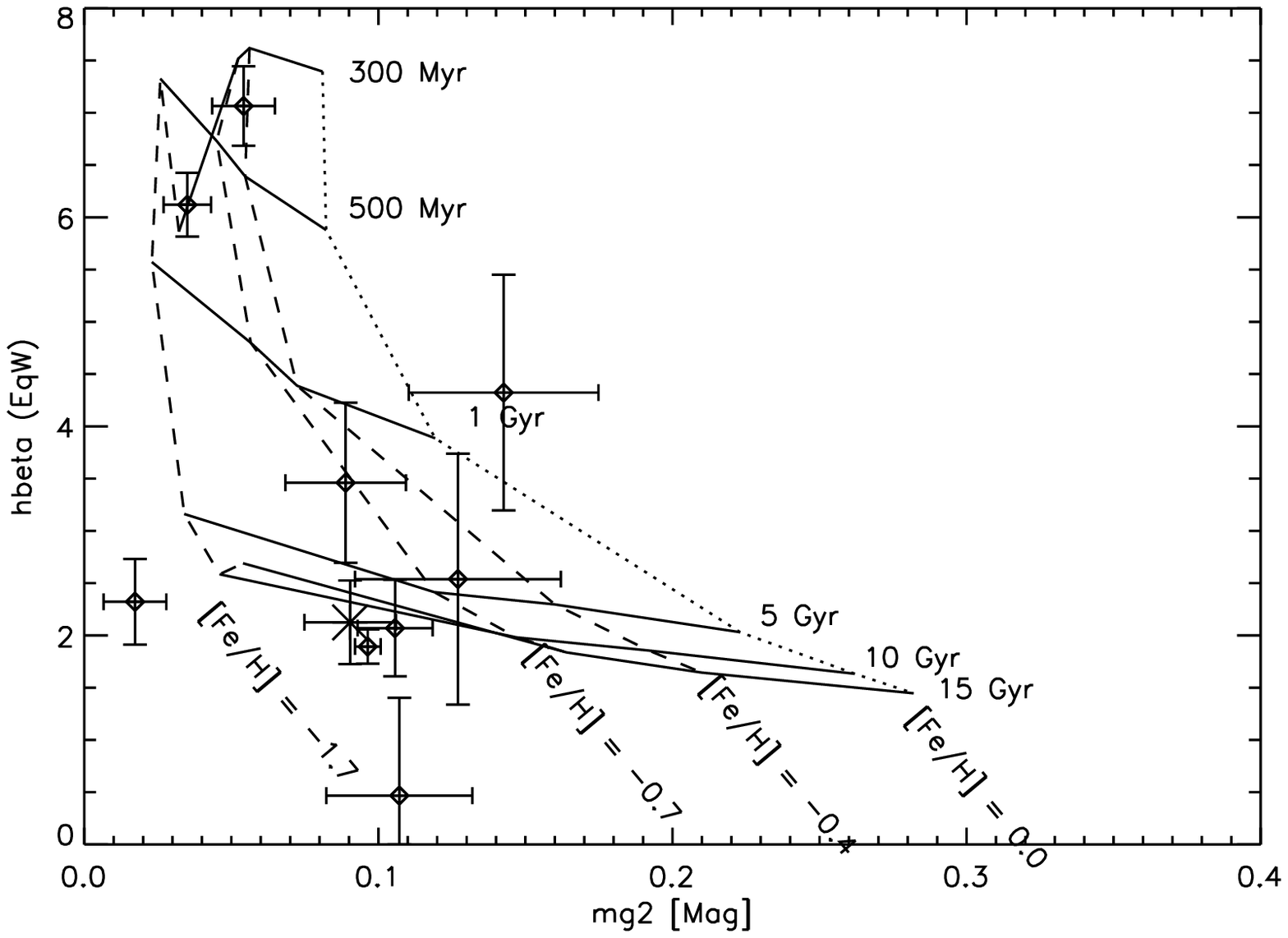}
\figcaption[Larsen.fig4.ps]{\label{fig:hb_mg2}Lick/IDS 
H$\beta$ vs.\ Mg2 indices, compared to \citet{bc2001} population 
synthesis models. The asterisk ($*$) marks the combined data for the
6 old metal-poor GCs.  }
\newpage

\epsfxsize=14cm
\epsfbox{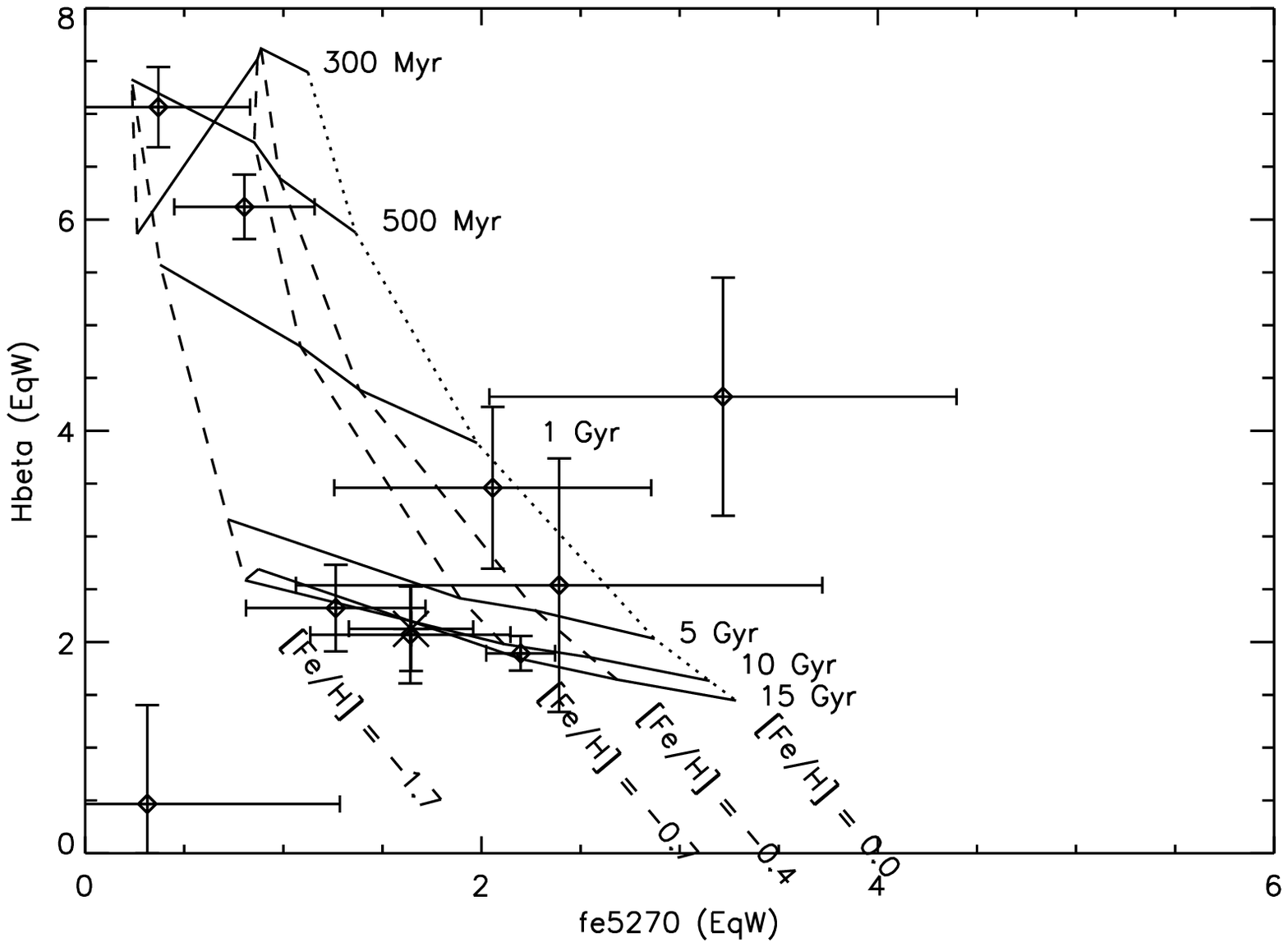}
\figcaption[Larsen.fig5.ps]{\label{fig:hb_fe5270}Lick/IDS 
H$\beta$ vs.\ Fe5270 indices, 
compared to \citet{bc2001} population synthesis models.}
\newpage

\epsfxsize=14cm
\epsfbox{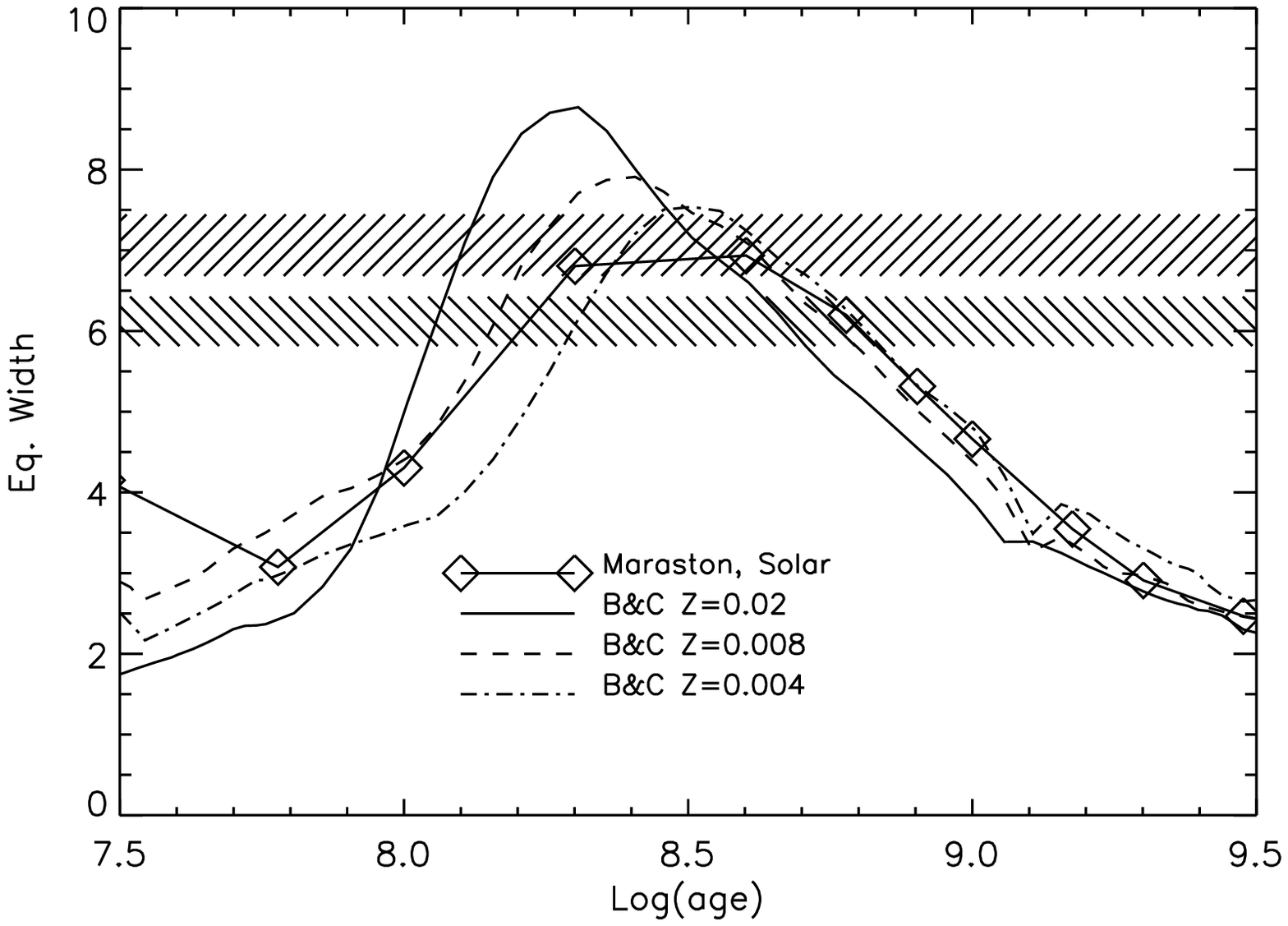}
\figcaption[Larsen.fig6.ps]{\label{fig:hb_cc_gw}
  Models for the Lick/IDS H$\beta$ index as a function of age, compared 
to the measured values for the two blue objects in NGC~1023A, n1023-3 and
n1023-4. The hatched areas indicate the measured values and their 
uncertainties.
}
\newpage

\epsfxsize=13cm
\epsfbox{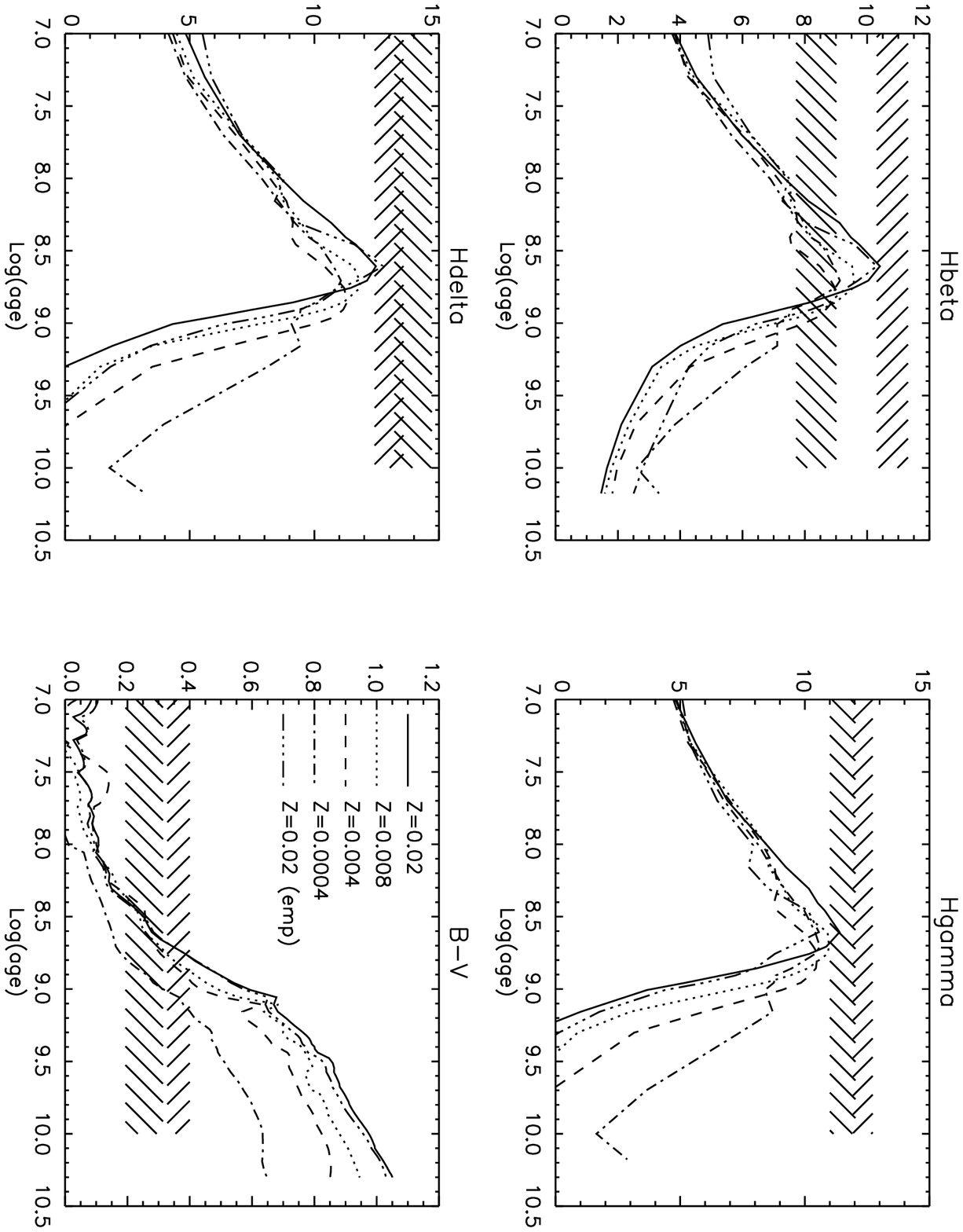}
\figcaption[Larsen.fig7.ps]{\label{fig:ccspec}
  Balmer line equivalent widths and \bv\ color for the two blue objects in 
NGC~1023A, compared to predictions based on Bruzual \& Charlot (2001)
models. The various curves in each diagram represent different 
metallicities.
}

\newpage

\begin{deluxetable}{llllrrrrr}
\tablecaption{ \label{tab:spec}
  Spectroscopic data for globular cluster candidates in NGC~1023}
\tablecomments{RV = heliocentric radial velocity, $V$ = visual magnitude. 
  The S/N column gives the average Signal-to-Noise in the region 
  4900\AA -- 5100 \AA . 
  $^*$ photometry from HST image in $0\farcs5$ aperture
}
\tablehead{Object   & Obs    & RA(2000)   & DEC(2000)  
           & $V$            & \vi           & RV (km/s)  & S/N}
\startdata
n1023-3  & Ground & 2:40:37.91 & 39:03:30.8 & $20.9\pm0.05$  & $0.72\pm0.08$ & $729\pm18$ & 21.8 \\
n1023-4  & Ground & 2:40:36.98 & 39:03:27.4 & $20.6\pm0.03$  & $0.70\pm0.04$ & $736\pm17$ & 28.1 \\
n1023-7  & Ground & 2:40:33.14 & 39:03:48.5 & $22.4\pm0.29$  & $1.13\pm0.42$ &  -         &  0.1 \\
n1023-8  & HST    & 2:40:32.27 & 39:04:26.3 & $22.07\pm0.01$ & $1.21\pm0.01$ & $628\pm37$ &  2.4 \\
n1023-9  & HST    & 2:40:31.29 & 39:03:33.0 & $22.02\pm0.01$ & $0.98\pm0.01$ & $369\pm53$ &  3.5 \\
n1023-10 & HST    & 2:40:30.58 & 39:03:33.2 & $21.91\pm0.01$ & $1.26\pm0.01$ & $812\pm32$ &  2.8 \\
n1023-11 & HST    & 2:40:30.57 & 39:04:39.0 & $22.22\pm0.01$ & $1.27\pm0.01$ & $584\pm195$ &  3.5 \\
n1023-12 & Ground & 2:40:29.69 & 39:04:21.3 & $21.4\pm0.25$  & $0.65\pm0.53$ &  -         &  0.0 \\
n1023-13 & Ground$^*$ & 2:40:27.84 & 39:04:40.2 & $19.39\pm0.01$  & $1.14\pm0.01$ & $353\pm17$ & 53.0 \\
n1023-14 & HST    & 2:40:24.61 & 39:04:29.8 & $22.00\pm0.01$ & $0.99\pm0.01$ & $722\pm22$ &  7.0 \\
n1023-15 & HST    & 2:40:23.95 & 39:04:48.7 & $21.59\pm0.01$ & $1.30\pm0.01$ &  -         & 11.2 \\
n1023-16 & HST    & 2:40:21.02 & 39:04:26.3 & $21.13\pm0.01$ & $0.99\pm0.01$ & $736\pm18$ & 19.0 \\
n1023-17 & HST    & 2:40:20.56 & 39:04:57.0 & $21.44\pm0.01$ & $1.01\pm0.01$ & $449\pm25$ & 11.4 \\
n1023-18 & Ground & 2:40:19.62 & 39:04:24.6 & $21.3\pm0.14$  & $1.14\pm0.20$ &  $z=2.15$  & 12.6 \\
n1023-19 & HST    & 2:40:18.96 & 39:04:32.9 & $21.83\pm0.01$ & $1.03\pm0.01$ & $603\pm17$ &  9.6 \\
n1023-20 & Ground & 2:40:17.57 & 39:03:58.4 & $22.4\pm0.64$  & $1.71\pm0.75$ &  -         &  2.5 \\
n1023-21 & HST    & 2:40:16.83 & 39:03:51.4 & $20.94\pm0.01$ & $0.99\pm0.01$ & $625\pm20$ & 20.5 \\
n1023-23 & Ground & 2:40:14.75 & 39:04:10.2 & $21.9\pm0.17$  & $0.53\pm0.39$ &  -         &  0.0 \\
n1023-24 & Ground & 2:40:13.39 & 39:03:45.7 & $21.2\pm0.08$  & $0.94\pm0.13$ & $-125\pm16$ & 15.3 \\
n1023-25 & Ground & 2:40:12.99 & 39:04:29.1 & $22.3\pm0.13$  & $1.24\pm0.17$ & $421\pm29$ &  7.3 \\
\enddata
\end{deluxetable}

\begin{deluxetable}{lrrrrrrrrr}
\rotate
\tablewidth{0pt}
\tabletypesize{\footnotesize}
\tablecaption{Metallicity estimates for globular clusters with spectra with
S/N$>$5.  Uncertainties have been estimated as described in BH90.
\label{tab:metal}
}
\tablehead{
Object   & \multicolumn{7}{c}{Individual [Fe/H] estimators} & 
	  [Fe/H] (sp)    &  [Fe/H] (phot) \\
         & Mg2 &  MgH  & Gband &  CNB &  Fe5270 &  CNR & H$+$K
}
\startdata
n1023-13 & $-1.20\pm 0.34$ & $-1.42\pm 0.48$ & $-0.73\pm 0.34$ & -- & $-0.56\pm 0.62$ & $-0.79\pm 0.46$ & -- & $-0.95\pm 0.18$ &  $-1.11\pm 0.03$ \\
n1023-14 & $-0.95\pm 0.49$ & $-1.53\pm 0.74$ & $-1.37\pm 0.87$ & -- & $-0.55\pm 1.00$ & $-0.61\pm 0.69$ & $-0.95\pm 1.06$ & $-1.00\pm 0.21$ &  $-1.60\pm 0.03$ \\
n1023-16 & $-1.05\pm 0.36$ & $-1.24\pm 0.52$ & $-1.31\pm 0.45$ & -- & $-0.98\pm 0.68$ & $-1.54\pm 0.49$ & $-1.35\pm 0.58$ & $-1.24\pm 0.10$ &  $-1.60\pm 0.03$ \\
n1023-17 & $-1.17\pm 0.40$ & $-2.22\pm 0.58$ & $-2.29\pm 0.72$ & -- & $-0.82\pm 0.77$ & $-1.37\pm 0.59$ & -- & $-1.55\pm 0.36$ &  $-1.53\pm 0.03$ \\
n1023-19 & $-1.09\pm 0.42$ & $-1.20\pm 0.63$ & $-2.46\pm 0.73$ & -- & $-2.11\pm 0.82$ & $-2.06\pm 0.57$ & $-1.11\pm 0.93$ & $-1.62\pm 0.32$ &  $-1.47\pm 0.03$ \\
n1023-21 & $-2.08\pm 0.35$ & $-1.31\pm 0.51$ & $-1.69\pm 0.41$ & $-1.76\pm 0.46$ & $-1.16\pm 0.66$ & $-1.58\pm 0.47$ & $-1.36\pm 0.53$ & $-1.61\pm 0.14$ &  $-1.60\pm 0.03$ \\
n1023-25 & $-0.80\pm 0.47$ & $-0.32\pm 0.72$ & $-0.96\pm 1.02$ & -- & $ 0.59\pm 0.95$ & $-1.86\pm 0.67$ & $-3.35\pm 1.07$ & $-1.05\pm 0.66$ &  $-0.78\pm 0.56$ \\
\enddata
\end{deluxetable}

\begin{deluxetable}{lrrr}
\tablecaption{\label{tab:ctest}Comparison of \vi\ colors from WFPC2
photometry and pseudo-colors measured on Keck spectra}
\tablehead{Objects  &  \vi\   &  $m_{4700} - m_{5700}$}
\startdata
n1023-3  & 0.72 & $-0.32$ \\
n1023-4  & 0.70 & $-0.30$ \\
n1023-13 & 1.14 & $ 0.11$ \\
n1023-14 & 0.99 & $-0.09$ \\
n1023-16 & 0.99 & $ 0.00$ \\
n1023-17 & 1.01 & $ 0.06$ \\
n1023-19 & 1.03 & $ 0.05$ \\
n1023-21 & 0.99 & $ 0.02$ \\
n1023-25 & 1.24 & $ 0.07$ \\
\enddata
\end{deluxetable}

\begin{deluxetable}{lrrrrr}
\rotate
\tablewidth{0pt}
\tabletypesize{\footnotesize}
\tablecaption{\label{tab:indices_bh}
Brodie \& Huchra indices 
}
\tablecomments{H$\gamma$ and H$\delta$ are defined as in 
Brodie \& Hanes (1986).}
\tablehead{
  Object & H$\beta$ (\AA) & H$\gamma$ (\AA) & H$\delta$ (\AA) & CNR & H$+$K
}
\startdata
n1023-13 & $0.084\pm0.007$ & $0.092\pm0.011$ & $0.030\pm0.014$ 
    & $0.054\pm0.011$ & -- \\
n1023-14 & $0.123\pm0.052$ & $0.099\pm0.076$ & $0.035\pm0.092$ 
    & $0.080\pm0.072$ & $0.286\pm0.123$ \\
n1023-16 & $0.088\pm0.019$ & $0.099\pm0.028$ & $0.125\pm0.037$ 
    & $-0.047\pm0.025$ & $0.237\pm0.050$ \\
n1023-17 & $0.148\pm0.034$ & $0.164\pm0.059$ & $0.016\pm0.069$ 
    & $-0.025\pm0.051$ & -- \\
n1023-19 & $0.003\pm0.037$ & $0.071\pm0.056$ & $0.164\pm0.075$ 
    & $-0.119\pm0.047$ & $0.267\pm0.105$ \\
n1023-21 & $0.100\pm0.017$ & $0.087\pm0.024$ & $0.104\pm0.029$ 
    & $-0.053\pm0.020$ & $0.235\pm0.039$ \\
n1023-25 & $0.183\pm0.052$ & $0.149\pm0.081$ & $0.124\pm0.094$ 
    & $-0.090\pm0.067$ & $-0.016\pm0.124$ \\
\enddata
\end{deluxetable}

\begin{deluxetable}{lrrrrrrrr}
\tablewidth{0pt}
\rotate
\tabletypesize{\footnotesize}
\tablecaption{\label{tab:stdcmp}Lick/IDS standard star observations}
\tablehead{
         & \multicolumn{4}{c}{HR3905} & \multicolumn{4}{c}{HR1805} \\
  Index  &  STD   &   Flux calibr.    &   Unfluxed        & Smoothed         
         &  STD   &   Flux calibr.    &   Unfluxed        & Smoothed
}
\startdata
Ca4227 &  1.820 & $ 2.151\pm 0.179$ &  $ 2.179\pm 0.177$ &  $ 1.961\pm 0.098$ &  2.450 & $ 3.005\pm 0.284$ &  $ 3.030\pm 0.275$ &  $ 2.680\pm 0.083$ \\
G4300  &  6.870 & $ 7.311\pm 0.127$ &  $ 7.415\pm 0.120$ &  $ 6.913\pm 0.061$ &  7.000 & $ 7.556\pm 0.212$ &  $ 7.679\pm 0.220$ &  $ 6.859\pm 0.113$ \\
hbeta  &  1.250 & $ 1.467\pm 0.064$ &  $ 1.453\pm 0.064$ &  $ 1.337\pm 0.060$ &  0.850 & $ 1.411\pm 0.364$ &  $ 1.397\pm 0.363$ &  $ 1.312\pm 0.310$ \\
MgB    &  4.290 & $ 4.518\pm 0.032$ &  $ 4.524\pm 0.032$ &  $ 4.416\pm 0.029$ &  3.880 & $ 4.097\pm 0.048$ &  $ 4.104\pm 0.050$ &  $ 3.964\pm 0.066$ \\
Fe5270 &  4.130 & $ 4.577\pm 0.047$ &  $ 4.580\pm 0.047$ &  $ 4.231\pm 0.041$ &  4.340 & $ 4.906\pm 0.092$ &  $ 4.908\pm 0.093$ &  $ 4.586\pm 0.007$ \\
Fe5335 &  3.750 & $ 4.519\pm 0.098$ &  $ 4.516\pm 0.097$ &  $ 3.979\pm 0.103$ &  3.910 & $ 4.623\pm 0.180$ &  $ 4.622\pm 0.179$ &  $ 4.080\pm 0.131$ \\
Mg1    &  0.172 & $ 0.168\pm 0.001$ &  $ 0.163\pm 0.002$ &  $ 0.164\pm 0.001$ &  0.214 & $ 0.199\pm 0.005$ &  $ 0.196\pm 0.005$ &  $ 0.194\pm 0.005$ \\
Mg2    &  0.327 & $ 0.320\pm 0.001$ &  $ 0.315\pm 0.001$ &  $ 0.316\pm 0.001$ &  0.361 & $ 0.347\pm 0.002$ &  $ 0.344\pm 0.002$ &  $ 0.343\pm 0.002$ \\
\enddata
\end{deluxetable}

\begin{deluxetable}{lrrrrrrr}
\rotate
\tablewidth{0pt}
\tabletypesize{\footnotesize}
\tablecaption{\label{tab:indices_gw}
Various Lick / IDS indices.
}
\tablecomments{Avg: Average of clusters n1023-13, n1023-14, n1023-16,
n1023-17, n1023-19 and n1023-21}
\tablehead{
  Object & H$\beta$ (\AA) & G4300 (\AA) & Ca4227 (\AA) & Fe5270 (\AA) & Fe5335 (\AA) & Mg2 (mag) & CN2 (mag)
}
\startdata
n1023-3 & $7.065\pm0.380$ & $-1.469\pm0.535$ & $0.488\pm0.270$ & $0.369\pm0.463$ & $0.480\pm0.574$ & $0.054\pm0.011$ & $-0.161\pm0.014$ \\
n1023-4 & $6.120\pm0.305$ & $-1.242\pm0.415$ & $-0.014\pm0.211$ & $0.803\pm0.354$ & $0.836\pm0.438$ & $0.035\pm0.008$ & $-0.135\pm0.011$ \\
n1023-13 & $1.893\pm0.164$ & $4.222\pm0.329$ & $1.017\pm0.192$ & $2.198\pm0.174$ & $2.007\pm0.207$ & $0.096\pm0.004$ & $0.079\pm0.012$ \\
n1023-14 & $2.538\pm1.201$ & $2.587\pm2.076$ & $0.090\pm1.265$ & $2.392\pm1.328$ & $0.683\pm1.691$ & $0.127\pm0.035$ & $0.068\pm0.075$ \\
n1023-16 & $2.070\pm0.462$ & $2.742\pm0.820$ & $0.866\pm0.451$ & $1.641\pm0.505$ & $1.452\pm0.625$ & $0.106\pm0.013$ & $-0.030\pm0.026$ \\
n1023-17 & $3.461\pm0.766$ & $0.697\pm1.732$ & $1.056\pm0.940$ & $2.057\pm0.799$ & $1.667\pm0.985$ & $0.089\pm0.020$ & $-0.009\pm0.054$ \\
n1023-19 & $0.467\pm0.937$ & $1.179\pm1.764$ & $1.444\pm0.909$ & $0.313\pm0.972$ & $-1.096\pm1.212$ & $0.107\pm0.025$ & $-0.036\pm0.050$ \\
n1023-21 & $2.321\pm0.410$ & $2.294\pm0.676$ & $0.716\pm0.371$ & $1.264\pm0.453$ & $1.252\pm0.556$ & $0.017\pm0.011$ & $-0.030\pm0.021$ \\
n1023-25 & $4.323\pm1.128$ & $2.874\pm2.416$ & $-0.528\pm1.391$ & $3.219\pm1.179$ & $3.583\pm1.415$ & $0.143\pm0.032$ & $-0.064\pm0.072$ \\
Avg & $2.125\pm0.400$ & $2.287\pm0.510$ & $0.865\pm0.184$ & $1.644\pm0.314$ & $0.994\pm0.455$ & $0.090\pm0.016$ & $0.007\pm0.021$ \\
\enddata
\end{deluxetable}

\end{document}